\documentclass[prl,
amsmath,amssymb,
twocolumn,
reprint,
superscriptaddress,
citeautoscript
]{revtex4-1}

\usepackage{amsmath}
\usepackage{graphicx}
\usepackage{graphics}
\usepackage{epstopdf}
\usepackage{float}
\usepackage[caption=false]{subfig}
\usepackage{color} 
\usepackage{tabularx}
\usepackage{multirow}

\usepackage{enumitem}

\begin{document}
\title{The cause of extremely long-lasting room-temperature persistent photoconductivity in SrTiO$_3$ and related materials}

\author{Zhiqiang Zhang}
\affiliation{Department of Physics and Astronomy, \\
University of Delaware, Newark, DE 19716, USA}
\affiliation{Department of Materials Science and Engineering, University of Delaware, Newark, DE 19716, USA}
\author{Anderson Janotti}
\email{janotti@udel.edu}
\affiliation{Department of Materials Science and Engineering, University of Delaware, Newark, DE 19716, USA}
\begin{abstract}

It has been recently revealed that strontium titanate (SrTiO$_3$) displays persistent photoconductivity with unique characteristics: it occurs at room temperature and lasts over a very long period of time. Illumination of SrTiO$_3$ crystals at room temperature with sub-bandgap light reduces the electrical resistance by three orders of magnitude and persists for weeks or longer (Tarun {\em et al.}, Phys. Rev. Lett. {\bf 111}, 187403 (2013)). Experiments indicate that oxygen vacancy and hydrogen play important roles, yet the microscopic mechanism responsible for this remarkable effect has remained unidentified. Using hybrid density functional theory calculations we show that an instability associated with substitutional hydrogen H$_{\rm O}^+$ under illumination, which becomes  doubly ionized and leaves the oxygen site, can explain the experimental observations. H$_{\rm O}$ then turns into an intersititial hydrogen and an oxygen vacancy, leading to excess carriers in the conduction band. This phenomenon is not exclusive to SrTiO$_3$, but it is also predicted to occur in other oxides. Interestingly, this phenomenon represents an elegant way of proving the existence of hydrogen substituting on an oxygen site (H$_{\rm O}$), forming an interesting, and rarely observed, type of three-center two-electron bond.

\end{abstract}
\date{\today}

\maketitle
Photoconductivity is a fundamental physical process in semiconductors where excess carriers created under illumination lead to often dramatic enhancement in electrical conductivity \cite{Queisser1985}. In general, the conductivity goes back to its original value after the illumination is switched off. Interesting cases happen when the photogenerated increased conductivity persists for very long times after the exciting light has been terminated. In these cases, the excess carriers remain delocalized in the bands, due to some microscopic mechanism that hinders carrier recombination.  Persistent photoconductivity (PPC) in semiconductors is typically observed at low temperatures \cite{Lang1977}, i.e. below about 77 K, meaning that recombination of the photo-generated carriers is obstructed by relatively low thermally activated energy barriers in real, configurational, or momentum spaces \cite{Queisser1985}.  Recent experiments \cite{Tarun2013} have shown that SrTiO$_3$, with indirect (direct) band gap of 3.25 eV (3.75 eV) \cite{Benthem2001}, exhibits extremely long living PPC at room temperature upon exposure to light of energy 2.9 eV or higher. The microscopic mechanism behind this extraordinary effect observed at {\em room temperature} remains unknown.

\begin{figure}[t!]
\includegraphics[width=3.2 in]{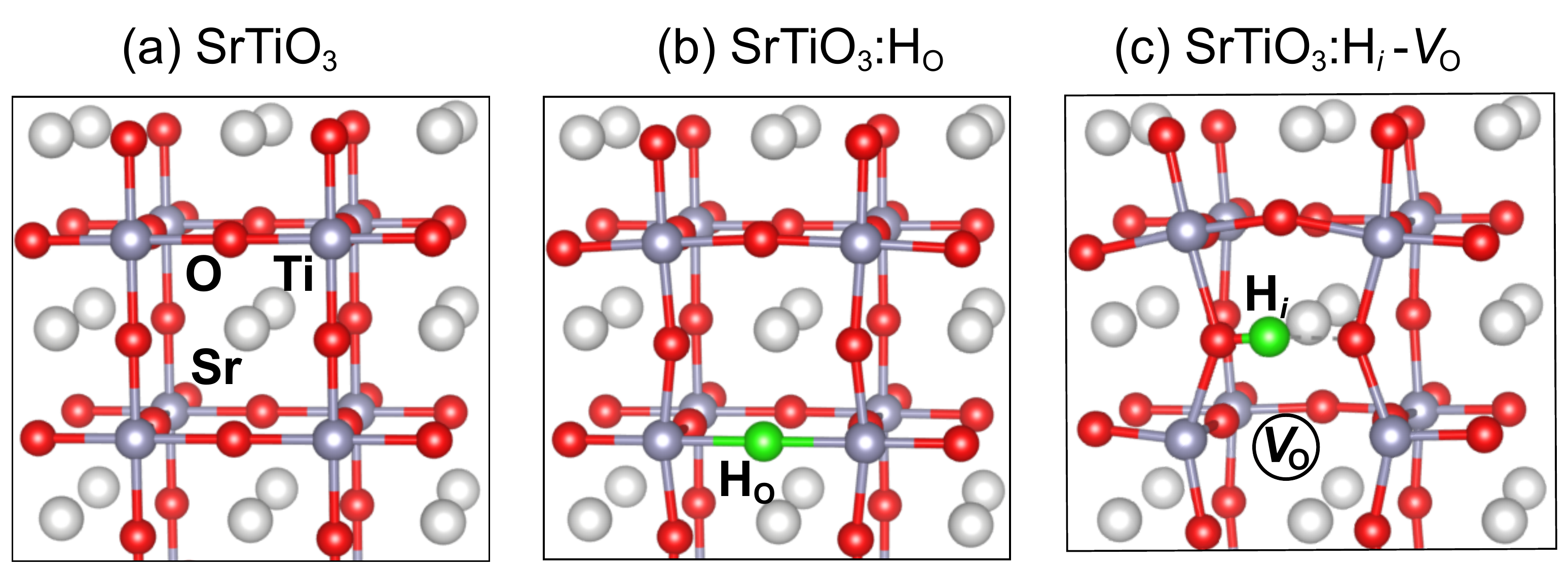}
\caption{\label{structures} Ball \& stick model of (a) bulk SrTiO$_3$, (b) substitutional H (H$_{\rm O}^+$) and (c) complex of oxygen vacancy ($V_{\rm O}^+$) and interstitial H (H$_i^+$) in SrTiO$_3$.}
\end{figure}

SrTiO$_3$ (STO) has the cubic perovskite crystal structure at room temperature (Fig.~\ref{structures}(a)) and is often used as substrate for the growth of other oxides, including superconducting thin films \cite{wu1987epitaxial}. The discovery of two-dimensional electron gas at the interface of STO with other complex oxides such as LaAlO$_3$ \cite{ohtomo2004high} and GdTiO$_3$ \cite{moetakef2011electrostatic} has fueled interest in its application in oxide electronics. Through doping, the electrical properties of STO can be tuned from insulating to semiconducting, metallic, and even superconducting \cite{Hosler1964,Chapman1967,Cohen1964,Janotti2012}. 
Experiments have shown that STO with specific annealing treatment exhibits PPC at room temperature upon exposure to sub-band gap light excitation\cite{Tarun2013,Poole2018}; the conductivity increases by three orders of magnitude and lasts over a year, making STO quite unique since sizeable PPC in semiconductors is normally observed only at low temperatures \cite{Lang1977}.
A tentative model for the mechanism behind PPC was proposed, in which hydrogen impurities seem to play critical role \cite{Poole2018}. 

Hydrogen is an ubiquitous impurity in oxides and can greatly alter their physical properties \cite{Vandewalle2000,Janotti2007,limpijumnong2009hydrogen,King2009shallow}, making them electrically conductive. Experiments have demonstrated that hydrogen impurities easily incorporate into STO \cite{Tarun2011}, with local vibrational mode frequencies in the range of 3350-3600 cm$^{-1}$, associated with the stretching of O-H bonds. These signals were attributed to interstitial hydrogen (H$_i$) bonded to an O atom either at a regular site or within a Sr or Ti vacancy based on 
density functional theory (DFT) and hybrid functional calculations \cite{Thienprasert2012,Varley2014}.
It was also predicted that hydrogen can incorporate on the oxygen site (H$_{\rm O}$) \cite{Varley2014}, as shown in Fig.~\ref{structures}(b), and that both H$_i$ and H$_{\rm O}$ act as shallow donors and can contribute to $n$-type conductivity. Yet a microscopic mechanism for the role of hydrogen in the observed PPC in STO has remained to be unveiled.

\begin{figure}[t!]
\includegraphics[width=2.9in]{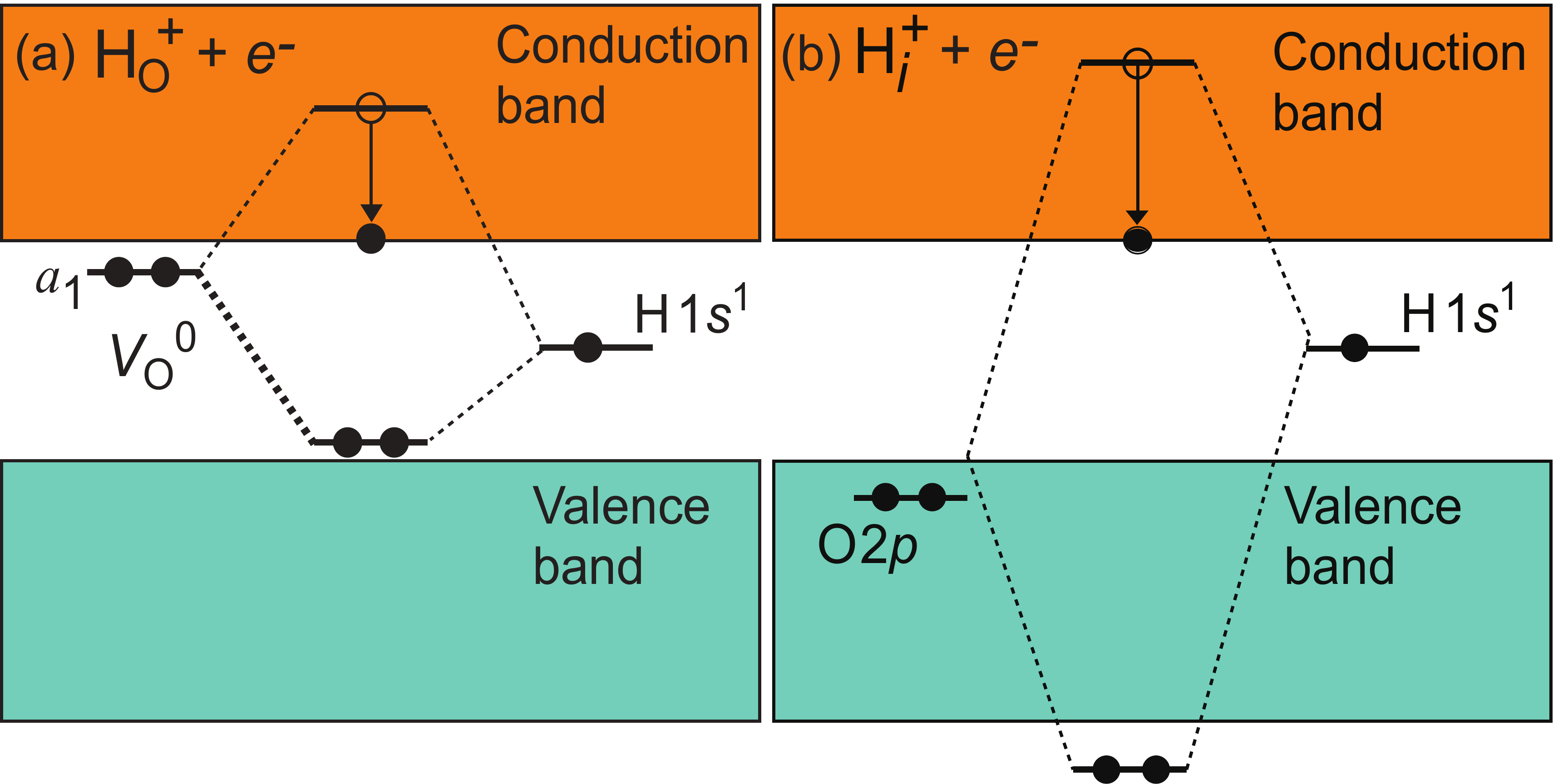}
\caption{\label{schematic} Schematic of the electronic structure of (a) H$_{\rm O}^+$ (plus an electron at the CBM) based on the combination of the neutral $a_1$  oxygen vacancy ($V_{\rm O}^0$) molecular orbital \cite{Janotti2014} and the H 1$s$ orbital and (b) H$_{\rm i}^+$ (plus an electron at the CBM) based on the combination of an oxygen 2$p$ orbital and the $a_1$ H 1$s$ orbital.
}
 \end{figure}

Here we use first-principles calculations based on DFT and hybrid functional to investigate the role of hydrogen in the observed PPC in STO. We find that H$_{\rm O}^+$ introduces a doubly occupied defect state near the valence-band maximum (VBM) of STO, as schematically shown in Fig.~\ref{schematic}(a).
Upon exposure to light with energy close to or higher than the band gap ($\hbar\omega \gtrsim 2.9$ eV), one electron from the H$_{\rm O}^+$-related state is excited to the conduction band,
transforming it into H$_{\rm O}^{2+}$, which in turn is metastable at room temperature. The H atom then moves out from the O site, bonding to an adjacent O in the form of H$_i^+$, 
leaving an ionized vacancy $V_{\rm O}$ behind, thus liberating electrons that contribute to the PPC. H$_i^+$ can easily diffuse in the material, eventually finding an Sr vacancy. We find that these results also apply to other oxides, such as TiO$_2$ and BaTiO$_3$, and that the observed PPC is an interesting way to probe the existence of substitutional H$_{\rm O}$ which is often difficult to detect in conducting oxides using standard vibration spectroscopic techniques \cite{Koch2012}.

Our calculations are based on DFT \cite{hohenberg1964inhomogeneous,kohn1965self} 
within the generalized gradient approximation (GGA) \cite{Perdew2008} 
and the HSE06 hybrid functional \cite{HSE1,HSE2}, and the projector augmented wave (PAW) potentials \cite{blochl1994projector,kresse1999ultrasoft} as implemented in the VASP code \cite{kresse1996efficiency,kresse1996efficient}. We use a plane-wave basis set with energy cutoff of 400 eV. The calculated indirect band gap of 3.22 eV is in good agreement with the experimental value of 3.25 eV for the cubic phase of STO at room temperature \cite{Benthem2001}. For the defect calculations we used a 135-atom supercell, which is a 3$\times$3$\times$3 repetition of the five-atom cubic primitive cell with calculated equilibrium lattice parameter of 3.913 \AA. Brillouin zone sampling was restricted to the $\Gamma$-point, yet tests using a 2$\times$2$\times$2 $k$-points mesh change formation energies by less than 0.1 eV. Defect formation energies were calculated as previously described \cite{Varley2014,Janotti2014,Freysoldt2014}. For charged defects, electrons are added or removed from the defect-related gap states, and a compensating background is added to keep the whole system charge neutral, and the formation energies are corrected due to the finite size of the supercell following the approach described in Ref.~[\onlinecite{Freysoldt2009}]. 
For the chemical potentials $\mu_{\rm O}$, $\mu_{\rm H}$, and $\mu_{\rm Sr}$, we used the values of Ref.~[\onlinecite{Varley2014}], representing experimental hydrogenation conditions. Note that all defect reaction barriers discussed here do not depend on the specific values of the chemical potentials since they conserve the number of atoms. Spin-polarization effects are included in all calculations with unpaired electrons.

\begin{figure}[t!]
\includegraphics[width=3.0in]{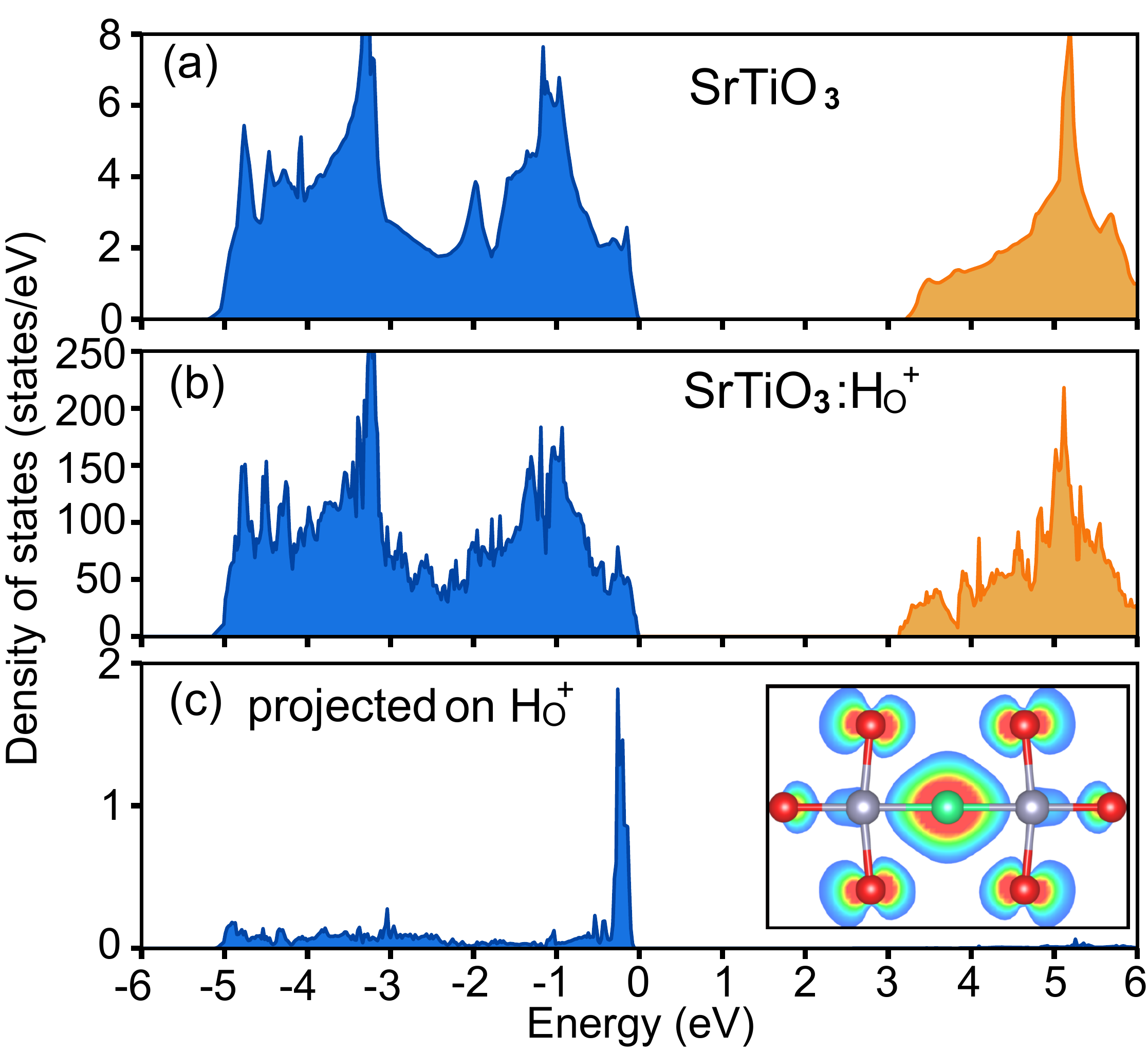}
\caption{\label{DOS} Total density of states of (a) SrTiO$_3$ for the 5-atoms bulk, (b) 135-atoms super cell with one  H$_{\rm O}^+$, and  (c) the orbital-projected density of states on H$_{\rm O}^+$.
}
 \end{figure}

Interstitial hydrogen can be bonded to a regular O atom, in which case it acts as a donor, H$_i^+$, or it can be bonded to an O atom neighboring a Sr or Ti vacancy \cite{Varley2014}, lowering the acceptor charge state of the vacancy. The electronic structure of H$_i^+$ can be understood as follows: a neutral H forms a chemical bond with an O, resulting in an doubly occupied bonding state several electron volts below the VBM,
and a singly occupied antibonding state well above the conduction-band minimum (CBM), which then lowers its energy by becoming a delocalized electron in the conduction band, as schematically shown in Fig.~\ref{schematic}(b).
The electronic structure of substitutional H$_{\rm O}^+$ can be described as follows: the H 1$s$ state combines with the doubly occupied symmetric $a_1$ state of the neutral O vacancy ($V_{\rm O}$), resulting in a bonding state near the VBM, and an antibonding state above the CBM. Two electrons occupy the bonding state and the third electron that would occupy the antibonding state is transferred to the conduction band, resulting in a shallow donor center H$_{\rm O}^+$, as schematically  drawn in Fig.~\ref{schematic}(a). This description is corroborated by inspecting the calculated orbital-projected density of states (DOS) of H$_{\rm O}^+$ in STO shown in Fig.~\ref{DOS}(c), where we compare the DOS of the perfect STO crystal and the supercell with one H$_{\rm O}^+$ defect in Fig.~\ref{DOS}(a) and (b). The calculated H$_{\rm O}$-Ti bond length is 2.0 \AA, compared with the equilibrium Ti-O bond length of 1.96 \AA~in STO.  
We note that H$_{\rm O}^+$ has been predicted and subsequently observed in ZnO \cite{Janotti2007,Koch2012}, and predicted to occur in other oxides such as SnO$_2$ \cite{Singh2008}, In$_2$O$_3$ \cite{limpijumnong2009hydrogen}, and TiO$_2$ \cite{Varley2011}. The observation of H$_{\rm O}^+$ in oxides other than ZnO has remained elusive.

In the experiments of Tarun {\em et al.} \cite{Tarun2013}, PPC occurs when the sample is exposed to light with photon energy of 2.9 eV or higher, which is slightly lower than the band gap of 3.25 eV. Our results show that H$_{\rm O}^+$ introduces a defect level near the VBM. One electron from this level can be lifted to the conduction band when exposed to light with photons of sub-band gap energy. 

\begin{figure}
\includegraphics[width=2.9 in]{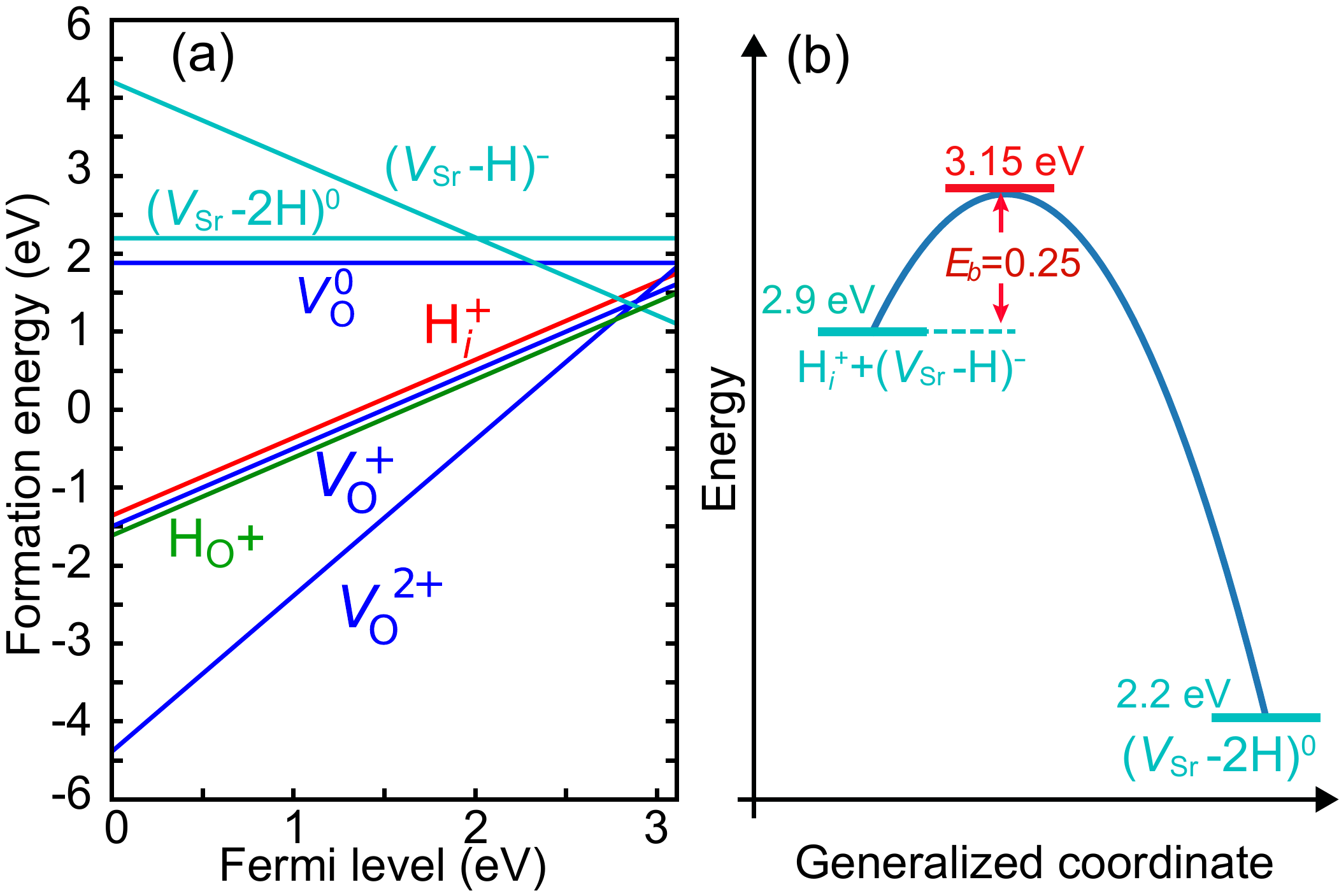}
\caption{
\label{Formation-energy}
(a) Calculated formation energies of $V_{\rm O}^0$, $V_{\rm O}^+$, $V_{\rm O}^{2+}$, 
${\rm H}_{\rm O}^+$, ${\rm H}_i^+$, $(V_{\rm Sr}{-}{\rm H})^-$ and $(V_{\rm Sr}{-}{\rm 2H})^0$ as a function of the Fermi level position. These results are essentially the same as those published in 
Ref.~[\onlinecite{Varley2014}], 
for the chemical potentials $\mu_{\rm O}$ = -3.2 eV and $\mu_{\rm H}$ = -0.61 eV, and $\mu_{\rm Sr}$ = -3.9 eV, i.e., point D in the phase diagram (Fig.~1) of 
Ref.~[\onlinecite{Varley2014}]. 
(b) Schematic representation of the energy barrier for the reaction  ${\rm H}_i^+ {+} (V_{\rm Sr}{-}{\rm H})^-$$\rightarrow$ $(V_{\rm Sr}{-}2{\rm H})^0$.
}
\end{figure}

\begin{figure}
\includegraphics[width=2.9 in]{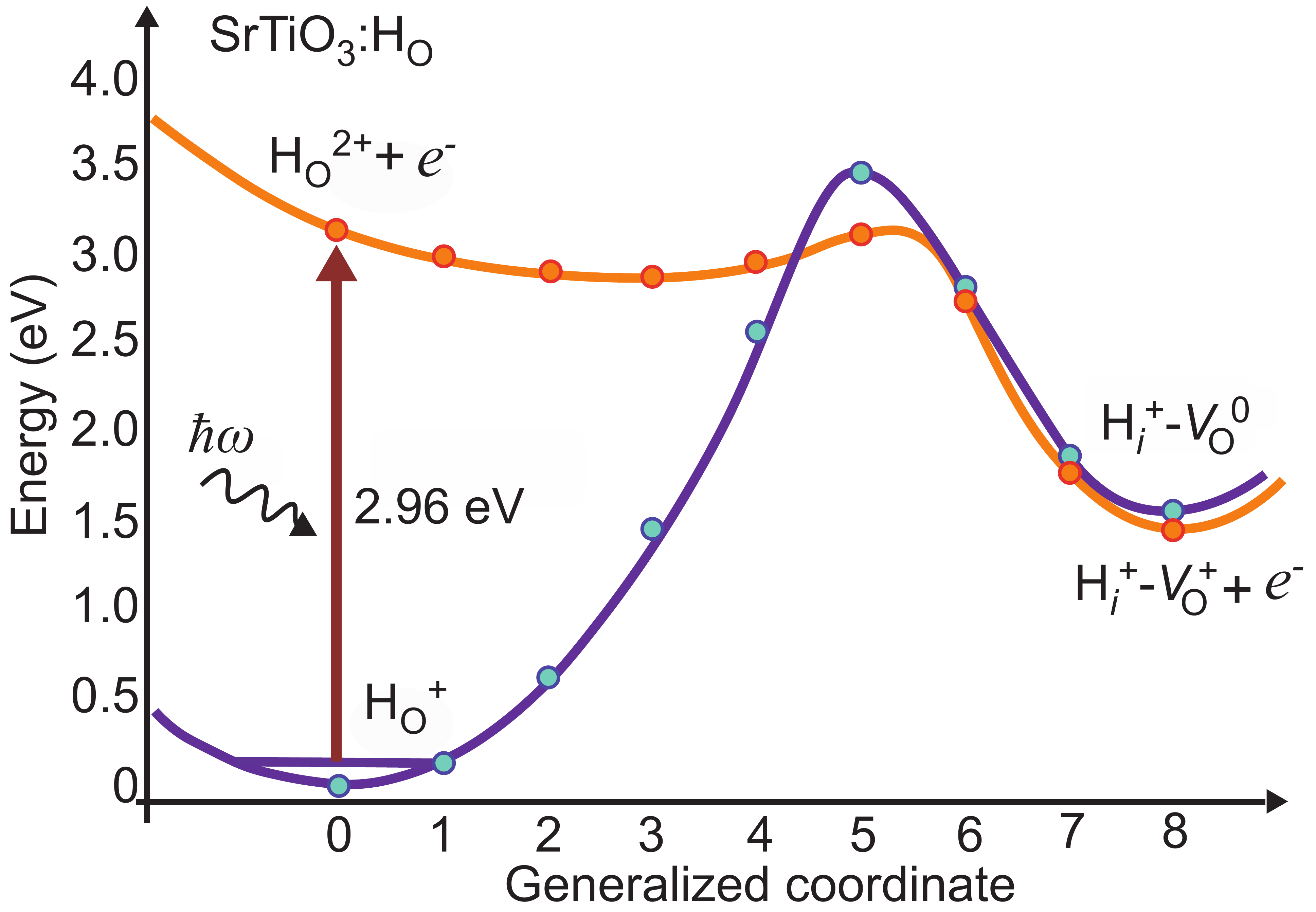}
\caption{\label{Barrier} Calculated energy barriers for the reactions Eq.~\ref{eq:1} and~\ref{eq:2} for the dissociation of ${\rm H}_{\rm O}^+$ (Fig. \ref{structures}(b)) into ${\rm H}_i^+$ and $V_{\rm O}^0$, and ${\rm H}_{\rm O}^{2+}$ into ${\rm H}_i^+$ and $V_{\rm O}^+$ (Fig.\ref{structures}(c)) in SrTiO$_3$. Note that we subtracted the zero point energy of ${\rm H}_{\rm O}^+$ (0.18 eV, determined from the calculated frequencies of the local vibration modes) in the excitation ${\rm H}_{\rm O}^++\hbar\omega \rightarrow {\rm H}_{\rm O}^{2+}+e^-$.
}
\end{figure}

Recent studies by Poole {\em et al.} \cite{Poole2018} reveal that the presence of hydrogen and oxygen during high temperature annealing plays a critical role in the observed PPC. Based on calculated defect formation energies and reaction barriers, shown in Fig.~\ref{Formation-energy} and Fig.~\ref{Barrier}, we propose the following model for the PPC mechanism:
\begin{equation} \label{eq:1}
{\rm H}_{\rm O}^+ + (V_{\rm Sr}{-}{\rm H})^-\rightarrow V_{\rm O}^{2+}+(V_{\rm Sr}{-}2{\rm H})^0+2e^{-1}.
\end{equation}
First, ${\rm H}_{\rm O}^+$ and $(V_{\rm Sr}{-}{\rm H})^-$defects, in which one H atom is bonded to an O in the Sr vacancy, are introduced during the high-temperature annealing process under the "hydrogen rich" condition, resulting in almost fully compensated samples with low electrical conductivity. Upon exposure to light, one electron from the ${\rm H}_{\rm O}^+$-related level is excited to the conduction band, resulting in ${\rm H}_{\rm O}^{2+}$:
\begin{equation} \label{eq:2}
{\rm H}_{\rm O}^++\hbar\omega \rightarrow {\rm H}_{\rm O}^{2+}+e^-.
\end{equation} 
${\rm H}_{\rm O}$ in the doubly positive charge state is metastable, and the H atom sees a barrier of only 
0.4 eV to leave the substitutional site and become an interstitial hydrogen (H$_i^+$), which is stabilized by forming a H-O bond with an nearby oxygen atom. This process is represented by the following charge-conserving equation:
 \begin{equation}\label{eq:3}
{\rm H}_{\rm O}^{2+} \rightarrow V_{\rm O}^{+}+{\rm H}_i^+;
 \end{equation}
the calculated energy barrier along the reaction path is shown in Fig.~\ref{Barrier}. The 7 intermediate configurations along the reaction path were generated by interpolating the atomic positions of the initial and final configurations as in the nudged elastic band approach \cite{NEB}. The final configuration, $V_{\rm O}^{+}+{\rm H}_i^+$, is shown in Fig.~\ref{structures}(c), in which the H binds to one O atom near the vacancy site and points to another O. The product $V_{\rm O}^{+}$ represents one small polaron bound to the doubly positive O vacancy, forming a singly positive complex \cite{Janotti2014}. 

The oxygen vacancy, in turn, is a shallow donor, with the (2+/+) transition level close to the conduction band, and the neutral charge state $V_{\rm O}^0$ is always higher in energy, even for the Fermi level at the CBM. Therefore, we expect that a large fraction of the oxygen vacancies will become doubly ionized $V_{\rm O}^{2+}$ at room temperature, contributing with additional free electrons to the conduction band, i.e., $V_{\rm O}^{+} \rightarrow V_{\rm O}^{2+}+e^-$.

Without exposure to light excitation, ${\rm H}_{\rm O}^+$ is quite stable at room temperature, and is unlikely to move out of the O site. The calculated energy barrier for the charge-conserving reaction ${\rm H}_{\rm O}^+ \rightarrow {\rm H}_{i}^{+}+V_{\rm O}^{0}$ is 3.5 eV, as shown in Fig.~\ref{Barrier}. Here $V_{\rm O}^{0}$ represents two small polarons bound to the doubly positive O vacancy $V_{\rm O}^{2+}$ forming a neutral complex \cite{Janotti2014}.
The barrier for ${\rm H}_{\rm O}^+$ to dissociate into ${\rm H}_{i}^{+}$ and a charged oxygen vacancy, $V_{\rm O}^{+}$ or $V_{\rm O}^{2+}$, with the release of one or two electrons for balancing the reaction, is also similarly high since the formation energies of the vacancy in the different charge states are quite close to each other when the Fermi level is at the CBM (see Fig.~\ref{Formation-energy}(a)). 

Once the H atom moves out of the O site, becoming ${\rm H}_i^+$, the barrier to come back to the O site, forming ${\rm H}_{\rm O}^{2+}$, is 1.6 eV as shown in Fig.~\ref{Barrier}, which is much higher than the barrier for ${\rm H}_i^+$ to move away from the O vacancy, bonding to another O that is further away from the vacancy.  This later energy barrier, we estimate, is just the migration barrier of ${\rm H}_i^+$ in STO, which is only 0.25 eV as calculated previously \cite{Thienprasert2012}. The Coulomb repulsion between ${\rm H}_i^+$ and $V_{\rm O}^+$ also favors to drive them apart, leading to an energy drop of 0.9 eV according to our calculations, which is the total energy difference between ${\rm H}_i^+$-$V_{\rm O}^+$ (Fig.~\ref{structures}(c)) and each defect separated.

We now turn to the second part of reaction in Eq.~\ref{eq:1}, i.e., the interaction between ${\rm H}_i^+$ and a Sr vacancy. As ${\rm H}_i^+$ diffuses in the STO, it will be attracted to negatively charged $(V_{\rm Sr}{-}{\rm H})^-$ complexes, assumed to be present after the annealing process described in \cite{Tarun2013,Poole2018}. The reaction 
\begin{equation}\label{eq:5} 
{\rm H}_i^+ + (V_{\rm Sr}{-}{\rm H})^- \rightarrow (V_{\rm Sr}{-}2{\rm H})^0
 \end{equation}
is exothermic and driven by the Coulomb attraction between the two oppositely charged centers, with an estimated barrier of only 0.25 eV (the same as the migration barrier of ${\rm H}_i^+$), with an energy gain of 0.7 eV for the system, as schematically shown in Fig.~\ref{Formation-energy}(b). This reaction does not involve release or capture of free electrons.  Once $(V_{\rm Sr}{-}2{\rm H})^0$ is formed, the barrier to dissociate into ${\rm H}_i^+ + (V_{\rm Sr}{-}{\rm H})^-$ is 0.95 eV.  Note that in $n$-type STO, i.e., with the Fermi level at the CBM, $(V_{\rm Sr}{-}2{\rm H})^0$ is more stable than separated ${\rm H}_i^+$ and $(V_{\rm Sr}{-}{\rm H})^-$ by 0.7 eV, meaning that the final products in Eq.~\ref{eq:1} are rather stable at room temperature and long-lived.


The reaction rate for processes in Eq.~\ref{eq:1} and~\ref{eq:2} can be estimated using 
\begin{equation}
\Gamma = \Gamma_0e^{-\frac{E_a}{k_BT}},
\end{equation}
where $E_a$ is the reaction barrier, $k_B$ is the Boltzman constant, $T$ is the temperature and $\Gamma_0$ can be approximated by the highest frequency of the local vibration mode (see below). If we take $\Gamma_0$=40~THz, $T$=300~K, we obtain $\Gamma=1.3\times10^{-45}~{\rm s}^{-1}$ for $E_a=3.5~{\rm eV}$, and $\Gamma=8\times10^6~{\rm s}^{-1}$ for $E_a=0.4~{\rm eV}$, i.e., ${\rm H}_{\rm O}^+$ will not move out of the O site at room temperature, while, upon illumination,  ${\rm H}_{\rm O}^+$ turns into ${\rm H}_{\rm O}^{+2}$, which is metastable and easily moves out of the O site.

To aid in the experimental identification of ${\rm H}_{\rm O}^+$ in STO before illumination, we calculated the frequencies of the local vibration modes associated with the H impurity. Infrared (IR) spectroscopy peaks of around 3500 cm$^{-1}$ observed in hydrogen-doped STO samples have been assigned to interstitial H \cite{weber1986spectroscopy,klauer1992local,Tarun2011}. This assignment is further confirmed by DFT calculations \cite{Varley2014}. Recent measurements \cite{Tarun2011} also find two additional peaks in the 3350-3390 cm$^{-1}$ range, yet the exact microscopic configuration of these peaks is still under debate. These three modes are all assigned to stretching of the O-H bonds. 

In the case of ${\rm H}_{\rm O}^+$, H is bonded to two Ti atoms with H-Ti distances of 2.0 \AA. We thus expect the frequencies of the ${\rm H}_{\rm O}^+$ local vibration modes to be much lower than those associated with O-H bond stretching.
For the ${\rm H}_{\rm O}^+$ local vibration modes we obtained a non-degenerate mode at 1300 cm$^{-1}$ and a twofold degenerate mode at 837 cm$^{-1}$. Although these are much higher and outside the phonon spectrum of STO, it may be difficult to probe them due to possible overlap with absorption due to free-carriers.

Having discussed the role of ${\rm H}_{\rm O}$ in the observed PPC in STO at room temperature, we then ask if this effect is exclusive to STO. We explored if ${\rm H}_{\rm O}$ would behave in the same way in other oxides with similar band structure of STO, such as BaTiO$_3$ and TiO$_2$. These oxides have band gaps of 3.20 eV \cite{Wemple1970} and 3.05 eV \cite{Pascual1977}, respectively, with upper valence bands derived mostly from O 2$p$ orbitals and lower conduction bands derived from Ti 3$d$ orbitals, similar to STO. The total DOS and the orbital projected DOS on ${\rm H}_{\rm O}$ in BaTiO$_3$ are shown in Fig.~S1 in the Supplemental Material \cite{SUP}. This result shows that the ${\rm H}_{\rm O}^+$-related state resides near the VBM in BaTiO$_3$, and electron from this state can be lifted to the conduction band using light with photons of near or higher energies than the band gap, as in STO.

The configuration coordinate diagram corresponding to the reaction path of Eqs.~\ref{eq:1} and~\ref{eq:2} for ${\rm H}_{\rm O}$ in BaTiO$_3$, displayed in Fig.~S2 in the Supplemental Material \cite{SUP}, also shows that while ${\rm H}_{\rm O}^+$ is stable at room temperature, with an energy barrier of 3.8 eV for dissociation into $V_{\rm O}^0$ and ${\rm H}_i^+$, upon illumination with photons of energy near or higher than the band gap, ${\rm H}_{\rm O}^+$ is then transformed into ${\rm H}_{\rm O}^{2+}$ which is metastable, with an energy barrier of only 0.3 eV to dissociate into $V_{\rm O}^+$ and ${\rm H}_i^+$, liberating electrons to the conduction band. 

For TiO$_2$, we find that ${\rm H}_{\rm O}^+$ sees an energy barrier of 2.4 eV for dissociation, whereas ${\rm H}_{\rm O}^{2+}$ sees a barrier of only 0.5 eV (not shown). Here too, we find that the ${\rm H}_{\rm O}^+$-related state is near the top of the valence band. Therefore, we expect that once ${\rm H}_{\rm O}$ is incorporated with concentrations that are much higher than the free electron concentration, due to compensation effects, we expect that illumination with photons with near band gap energies or higher will lead to long lasting PPC at room temperature.

In summary, our calculations show that hydrogen incorporated on oxygen site (${\rm H}_{\rm O}^+$) can explain the observed long lasting PPC in STO.  ${\rm H}_{\rm O}^+$ is initially very stable at room temperature, but upon exposure to light with near band gap energy, one electron from the ${\rm H}_{\rm O}^+$-related state is lifted to the conduction band; the resulting  ${\rm H}_{\rm O}^{2+}$ is metastable and leaves the O site, transforming into $V_{\rm O}^+$ and ${\rm H}_i^+$ with a barrier of only 0.4 eV. The reverse reaction is hindered by a larger barrier due to the Coulomb repulsion between the positively charged $V_{\rm O}^+$ and ${\rm H}_i^+$, which drives them apart. This effect is not exclusive to STO, but is also predicted to occur in BaTiO$_3$ and TiO$_2$ for example.

\section*{Acknowlegments}

We thank M. McCluskey and C.P. Pansegrau for fruitful discussions. This work was supported by the NSF Early Career Award grant number DMR-1652994, the Extreme Science and Engineering Discovery Environment (XSEDE) supported by National Science Foundation grant number ACI-1053575, and the Information Technologies (IT) resources at the University of Delaware.

\end{document}